\documentclass[preprint]{aastex}

\input{psfig.sty}
\def\etal   {{\it et~al.\/}}

\def\HII    {H~{\rm {II}}}

\def\oii{{[O\thinspace II]}}
\def\oiii{{[O\thinspace III]}}
\def\hb{{H\beta}}
\def\ew#1{W_{#1}}

\begin{document}

\title{Measuring Global Galaxy Metallicities Using Emission Line 
Equivalent Widths}

\author{Henry A. Kobulnicky}
\affil{Department of Physics \& Astronomy \\
University of Wyoming \\ Laramie, WY 82071 
\\ Electronic Mail: chipk@uwyo.edu}

\author{ Andrew C. Phillips}
\affil{University of California, Santa Cruz \\ Lick Observatory/Board
of Studies in Astronomy \\ Santa Cruz, CA, 95064  \\ 
phillips@ucolick.org}

;\author{Final Revised draft of 27 August 2003 }

\author{Accepted for Publication December 2003 in {\it The Astrophysical Journal} }

\vskip 1.cm

\begin{abstract}

We develop a prescription for estimating the interstellar
medium oxygen abundances of distant star-forming galaxies using the
ratio $EWR_{23}$ formed from the equivalent widths of the [O~II]
$\lambda3727$, [O~III] $\lambda\lambda4959,5007$ and H$\beta$ nebular
emission lines.  This $EWR_{23}$ approach essentially identical to the
widely-used $R_{23}$ method of Pagel \etal\ (1979).  Using data from
three spectroscopic surveys of nearby galaxies, we conclude that the
emission line equivalent width ratios are a good substitute for
emission line flux ratios in galaxies with active star formation.  The
RMS dispersion between $EWR_{23}$ and the reddening-corrected $R_{23}$
values is $\sigma(\log~R_{23})\leq0.08$ dex.  This dispersion is
comparable to the emission-line measurement uncertainties for distant
galaxies in many spectroscopic galaxy surveys, and is somewhat smaller than the
uncertainties of $\sigma(O/H)\sim0.15$ dex inherent in strong-line
metallicity calibrations.  Because equivalent width ratios are partially
insensitive to interstellar reddening effects, emission line
equivalent width ratios should be superior to flux ratios when
reddening corrections are not available. The $EWR_{23}$ method
presented here is likely to be most useful for statistically
estimating the mean metallicities for large samples of galaxies to
trace their chemical properties as a function of redshift or
environment.  The approach developed here is used in a companion paper
(Kobulnicky \etal\ 2003) to measure the metallicities of star-forming
galaxies at $z=0.2 - 0.8$ in the Deep Extragalactic Evolutionary Probe
spectroscopic survey of the Groth Strip.

\end{abstract}


\section{Introduction}

Abundances of chemical elements in galaxies are
commonly measured using the emission lines emitted by astrophysical
nebulae (e.g., see Osterbrock 1989).  Recombination lines of hydrogen
(e.g., the Balmer series) or helium and forbidden lines of singly and
doubly ionized carbon, nitrogen, oxygen, neon, and sulfur are among
the strongest observable lines in the visible and ultraviolet
spectrum.  The relative strengths of each emission line, combined with
some knowledge of the temperature and density of the nebulae, provide
information about the relative and absolute abundances of each ion.
Chemical analyses of individual nebulae have been used to test
nucleosynthesis models of stars, the chemical evolutionary history of
galaxies, and nucleosynthesis in the big bang (review by Aller 1990;
Pagel 1998).  Even integrated spectra of entire galaxies may be used
to estimate the overall degree of chemical enrichment of a
galaxy (Kobulnicky, Kennicutt, \& Pizagno 1998).

Observational techniques sometimes permit the relative fluxes of
emission lines from galaxies to be measured with a high degree of
precision.  However, in the current generation of wide-field galaxy
surveys on multi-object spectrographs, flux calibration is frequently
problematic due to unfavorable observing conditions or instrumental
effects such as a variation in system response over the field of view.
In addition, the traditional line ratio for reddening determination
is $H\alpha / H\beta$, but for redshifts greater than $z \sim 0.3$ the
$H\alpha$ line is not available, and these galaxies rarely have sufficient
signal-to-noise in their spectra to attempt a reddening derivation 
from the higher-order Balmer lines.
For such surveys, it is still desirable to extract chemical
information from the data, where possible.

In this paper we explore
the possibility of estimating mean gas-phase oxygen abundances for
galaxies based the {\it equivalent widths} (rather than fluxes) of
strong Balmer $H\beta$ recombination line and forbidden [O~II] and
[O~III] emission lines.
This method essentially uses the underlying stellar
continuum as a crude flux calibrator, and has the advantage that any
reddening affecting the stars and gas equally will be automatically removed.
While we would not expect the equivalent-width method to be as accurate
as the traditional method, particularly for individual objects, we find that it
works sufficiently well to enable us to measure approximate oxygen abundances
in relatively faint, intermediate-redshift galaxies.

Our approach is to develop the method from
basic principles, and to test and empirically calibrate it using three
spectroscopic galaxy datasets, and in the end compare the oxygen
abundance indicator $R_{23}$ to the quantity $EWR_{23}$ derived from
traditional line fluxes and flux ratios to results using only the
equivalent widths and equivalent width ratios.
The thesis of this
paper is that even when the dereddened line {\it flux} ratio $R_{23}$ is not
available, a corresponding ratio of {\it equivalent widths} can still
provide an estimate of the oxygen abundance.

This new method is applied to the study of abundances in a
set of intermediate-redshift galaxies observed as part of
the Deep Extragalactic Evolutionary Probe (DEEP) spectroscopic
survey in the Groth Strip (DGSS) presented in a companion paper,
Kobulnicky \etal\ (2003; Ke03).

\section{Gas-phase Oxygen Abundance Measurements from Emission Lines} 

Standard techniques for measuring chemical abundances in astrophysical
nebulae from the fluxes and flux ratios of nebular emission lines are
reviewed in Osterbrock (1989).  The most direct and reliable
techniques involve measuring a suite of temperature-sensitive and
density-sensitive line ratios to determine the physical conditions of
the plasma.  In the absence of high signal-to-noise data measuring
temperature--sensitive line ratios (e.g.,
[O~III]~$\lambda$4363/[O~III]~$\lambda$5007), the default diagnostic
for measuring the oxygen abundance of ionized nebulae has become the
ratio of strong oxygen emission lines $R_{23}\equiv
(I_{3727}+I_{4959}+I_{5007})/I_{H\beta}$ (Pagel \etal\ 1979).
Subsequently, many authors have developed formulations relating this
strong line ratio to the gas-phase oxygen abundance (e.g., Edmunds \&
Pagel 1984; McCall, Rybski \& Shields 1985; Dopita \& Evans 1986;
McGaugh 1991).  Most modern calibrations relating $R_{23}$ to oxygen
abundance include a measure of the ionization parameter, such as
$O_{32}\equiv (I_{4959}+I_{5007})/I_{3727} \equiv I_{\oiii}/I_{\oii}$
as a second parameter (e.g., McGaugh 1991; Pilyugin 2001).

While the $R_{23}$ method was developed primarily for individual HII
regions, Kobulnicky, Kennicutt, \& Pizagno (1998; KKP) have shown that
this approach is valid even when applied to the integrated spectra of
galaxies as a means of estimating an average O/H abundance for the
galaxy as a whole.  Even when the signal to noise ratio of these lines
is as low as 8:1 or when a spectroscopic observation encompasses a
range of metallicities and ionization conditions within a galaxy, this
ratio provides a rough, but reliable, estimate of the mean gas-phase
oxygen abundance when used in conjunction with the appropriate
calibration relating $R_{23}$ to $O/H$.  At this signal-to-noise
ratio, the associated uncertainty on the oxygen abundance due to line
measurement error is typically comparable to the uncertainty due to
the calibration of the $R_{23}$ vs. O/H relationship: $\sim0.15$ dex.
In this paper we show that the uncertainty introduced by substituting
emission line equivalent widths is less than these other sources of
uncertainty, and thereby establish a new method for measuring
interstellar medium oxygen abundances.

\section{Justification of the Equivalent Width Method}

The original $R_{23}$ method involves line ratios, so we consider the
effects of using equivalent width ratios in place of flux ratios.  For
generality, we include the effects of reddening in the discussion,
assuming the traditional ``screen'' model.
The equivalent width\footnote{We adopt the sign convention that emission line
equivalent widths are positive.} can be written as
\begin{equation}
 \ew{\lambda} = \frac{F_{\lambda 1}} {F_{C\lambda 1}}, 
\end{equation}
where $F_{\lambda}$ represents the line flux and $F_{C\lambda}$ is the
underlying continuum flux.  The fluxes do not need to be calibrated, since
the calibration affects both quantities equally.
The dereddened, calibrated flux value for a line, $I_\lambda$,
or continuum, $F_\lambda^0$, is given by
\begin{equation}
 I_{\lambda} = F^0_{\lambda} = F_{\lambda} \times 10^{c(1+f(\lambda))},
\end{equation}
where $c$ is the logarithmic attenuation at $H\beta$ and $f(\lambda)$
is a function describing the reddening curve (see Seaton 1979).
Combining these leads to a general line ratio formulation of
\begin{equation}
 \frac {I_{\lambda 1}} {I_{\lambda 2}} = 
   \frac {\ew{\lambda 1}} {\ew{\lambda 2}} \frac {F_{C\lambda 1}} {F_{C\lambda 2}}
    10^{c(f(\lambda 1) - f(\lambda 2))} =
    \frac {\ew{\lambda 1}} {\ew{\lambda 2}} \alpha,
\end{equation}
where we group all the continuum and reddening terms into the factor $\alpha$.
Note that the continuum will be attenuated by an amount characterized by
$c^*$, and reddened by (presumably) the same reddening law.  Thus,
\begin{equation}
 \alpha = \frac {F^0_{C\lambda 1}} {F^0_{C\lambda 2}} 
                 10^{(c - c^*)(f(\lambda 1) - f(\lambda 2))}.     
\end{equation}

The factor $\alpha$ contains two unknowns, the ratio of the non-reddened
continuum fluxes, which depends on the underlying stellar population,
and the difference in the attenuation of the emitting gas and the
continuum light.  Note that values of $c$ can range from zero to quite
large for individual HII regions, although large values are severely
deweighted in the average over an entire galaxy.  Typical derived
values are in the range 0 to 1, with a median value around 0.3 (McCall
\etal\ 1985; Olofsson 1995).  Values of $c^*$ can be estimated from
the relation $c^* = 1.47 E_{B-V}$ from Seaton.  For $E_{B-V}$ in a
typical range of 0--0.2, $c^* \sim$ 0--0.3. In physically realistic situations,
we would expect some correlation of $c$ and $c^*$, and also that ($c-c^*$) $\geq$ 0.

Let us now consider the specific line ratios of interest for the
$R_{23}$ method.  The first is $I_{\oiii} / I_{H\beta}$.  For the
purposes of reddening correction, we adopt the wavelength of the
stronger [O~III] $\lambda$5007 line.  The value of $\alpha$ in this
case is

\begin{equation}
\alpha_{\beta 3} = \frac {F^0_{C5007}} {F^0_{C4861}} {10^{(c - c^*)(-0.034)}}.
\end{equation}

Due to the proximity of these emission lines in wavelength, we expect
the flux ratios of the underlying stellar population to be very close to
unity. Similarly, for realistic $0 < (c-c^*) < 1$, the reddening factor
will be between unity and 0.92.  Thus, for this line ratio,
$\alpha \simeq 1$, and we will ignore it in further discussion.  
To a very good approximation,
\begin{equation}
 \frac {I_{\oiii}} {I_{\hb}} \simeq \frac {\ew{\oiii}} {\ew{\hb}} .
\end{equation}

The second ratio, $I_{\oii} / I_{\hb}$, is more problematic; for these
lines,
\begin{equation}
 \alpha_{2\beta} = \frac {F^0_{C3727}} {F^0_{C4861}} 10^{(c - c^*)(0.255)}.
\end{equation}
In principle, we could estimate $\alpha_{2\beta}$ from a detailed spectral
analysis of the flux-calibrated, integrated stellar spectrum and
Balmer line ratios, but for faint, galaxies with redshifts $z > 0.3$
we will have neither the signal-to-noise nor access to the H$\alpha$
line in order to do this.  In practice, we need to adopt an appropriate
average value of $\alpha_{2\beta}$ determined by other means.

Galaxy light tends to be dominated by A main sequence and G and K
giants (e.g., Morgan \& Mayall 1957;
Pritchet 1977; Kobulnicky \& Gebhardt 2000), and assuming an underlying
stellar spectrum composed of a linear combination of these two
spectral types implies that the dereddened flux ratio of the continuum
ranges from $\sim$1 for late-B stars to $\sim$0.4 for mid-G giants to
$\sim$0.2 for early-K giants (although galaxies whose $\lambda$3727
flux is dominated by K stars is unlikely to have any line emission
from star formation).  Examining the spectra in Kennicutt (1992b)
shows the $\lambda$3727-to-$\lambda$4861 ratio of the continuum fluxes
in galaxies with obvious emission lines ranging from $\sim$0.4--1.0;
if these were dereddened the range would shift upwards and might
possibly narrow somewhat.

On the other hand, the reddening correction ranges from 1 to 1.8 for
realistic values $0 \leq (c - c^*) \leq 1$, with a typical likely difference
$(c - c^*) \sim 0.3$ giving a reddening factor of $\sim$1.2.  Combining the
reddening and stellar population factors leads to an expected value of
$\alpha_{2\beta} \sim 0.84 \pm 0.3$.
This expected average value for $\alpha_{2\beta}$ differs from unity by less
than 0.1 dex.

We find the $R_{23}$ measure can now be expressed
\begin{equation}
R_{23} = \log \frac {I_{\oii} + I_{\oiii}} {I_{\hb}}
	  = \log \frac {\alpha_{2\beta} \ew{\oii} + \ew{\oiii}} {\ew{\hb}}.
\end{equation}
Not surprisingly, the $R_{23}$ measured using equivalent widths and an adopted
average value for $\alpha_{2\beta}$ will be most in error when [O~II] dominates
 [O~III].

The final line ratio of interest is the ionization parameter, and it is easy
to see that
\begin{equation}
  \frac{I_{\oiii}} {I_{\oii}} \simeq \frac{\ew{\oiii}} {\alpha_{23} \ew{\oii}}.
  \label{EQionpar}
\end{equation}

The value $\alpha_{23} \simeq \alpha_{2\beta}$ to a very good
approximation, so at this point we will drop the subscripts and simply
refer to ``$\alpha$'' for either of these quantities.  Comparing the
ratios above is probably the easiest means to empirically estimate an
average value for $\alpha$.  While it would be equally valid to
estimate $\alpha$ from the [O~II] and H$\beta$ ratios, the H$\beta$
equivalent width can be contaminated by underlying stellar absorption.


\section{Data Selection and Analysis}

In order to empirically measure an average value of $\alpha$ and to
confirm the validity of using equivalent width line ratios in
estimating $R_{23}$, we compiled three sets of
emission-line spectra of nearby galaxies.  The first
set consists of 16 objects with spatially-integrated spectra
from the 55-object spectroscopic galaxy
atlas of Kennicutt (1992a,b; K92) plus six additional emission-line
objects from KKP.  We refer to this sample as the K92+ sample.  The
K92+ spectra are produced by drifting a longslit across each galaxy
and have spectral resolutions of 5-7 \AA\ (K92) and 3 \AA\ (KKP).  The
16 K92 galaxies in our subsample are the strongest emission-line
objects where global metallicity measurements are possible.  The full
set of K92 galaxies represents a range of morphological types from Sa
to Im, but it includes only the bright galaxies of each type.  The KKP
galaxies are all underluminous, dwarf emission-line galaxies.  For a
second local sample, we examined the 198-galaxy Nearby Field Galaxy
Survey (Jansen \etal\ 2000a,b; NFGS), which is selected from the CfA
redshift catalog (Huchra \etal\ 1983).  From this spatially-integrated
spectroscopic survey, we culled 98 objects with
measurable $H\beta$, [O~III] and [O~II] emission lines.
These spectra have a resolution of 6 \AA\ and include a larger range
of luminosity ($-14<M_B<-22$) and surface brightnesses than K92+ while
spanning morphological types.  The K92+ objects have a higher fraction
of star-forming galaxies (objects with strong emission lines) compared
to the NFGS sample.  As a third local sample, we used emission-line
selected galaxies in the Kitt Peak National Observatory Spectroscopic
Survey (KISS; Salzer
\etal\ 2000).  KISS is a large-area objective prism survey of local
($z<0.09$) galaxies selected by strong $H\alpha$ emission, and thus,
preferentially contains objects with active star formation.  While
high-quality slit spectroscopy has been obtained for $\sim$519 KISS
galaxies (Melbourne \& Salzer 2002), 396 galaxies lack [O~II]
measurements, leaving 123 galaxies.  These remaining spectra have 5--8
\AA\ resolution, and are analyzed in Salzer \etal\ (2003). 

In summary, the 22 K92+, 98 NFGS, and 123 KISS galaxies were selected
from their larger parent samples because of strong emission lines
suitable for nebular metallicity measurement.  Following KKP, we chose
only galaxies with detectable [O~II]$\lambda$3727,
[O~III]$\lambda$4959, [O~III]$\lambda$5007, and H$\beta$ emission
lines.  Only galaxies where all four emission lines were measured with
a S/N of 8:1 or greater were retained.  This selection criterion
preferentially includes galaxies with high equivalent width lines, but
it also includes galaxies with low equivalent width lines where the
continuum is smooth and well-measured.

For each galaxy in the K92+ sample we measured the emission-line
fluxes and equivalent widths manually using Gaussian fits.  For the
NFGS and KISS surveys, we adopted the published emission-line fluxes
and equivalent widths.  Dereddened emission-line fluxes were computed
for all three samples by comparing the observed
$F_{H\alpha}/F_{H\beta}$ ratios to theoretical ratios.\footnote{
$I_{H\alpha}/I_{H\beta}$ = 2.75-2.86 for wide temperature range, e.g.,
(Hummer \& Storey 1987).  Here we assumed fixed electron temperature
of 12,000 K so that $I_{H\alpha}/I_{H\beta}$ = 2.85.}  The line fluxes
are dereddened using the law of Seaton (1979) as parameterized by
Howarth (1983) and as described in Kobulnicky \& Skillman (1996).  We
did not correct the Balmer emission lines for underlying stellar
absorption. The effects of Balmer absorption by the stellar population
are discussed in Section~5.3

While the K92+KKP sample consists entirely of starforming galaxies,
several low-level AGN are known to exist in the NFGS.  These four
objects are not included in the our subsample.  The 123 KISS galaxies
included here do not contain any AGN, as they were selected to be
conventional starforming galaxies based on analysis high-quality
spectroscopic observations (Salzer 2003).  In any case, the presence
of AGN among the samples would not have a significant bearing on the
results of this paper since we are interested in comparing observable
properties of emission lines rather than deriving physical quantities
such as density or metallicity which are sensitive to the nature of
the ionizing source.

\section{Analysis of Emission Line Quantities}

\subsection{[O~II] and [O~III] Equivalent Widths versus Fluxes}

\begin{figure}
\plotone{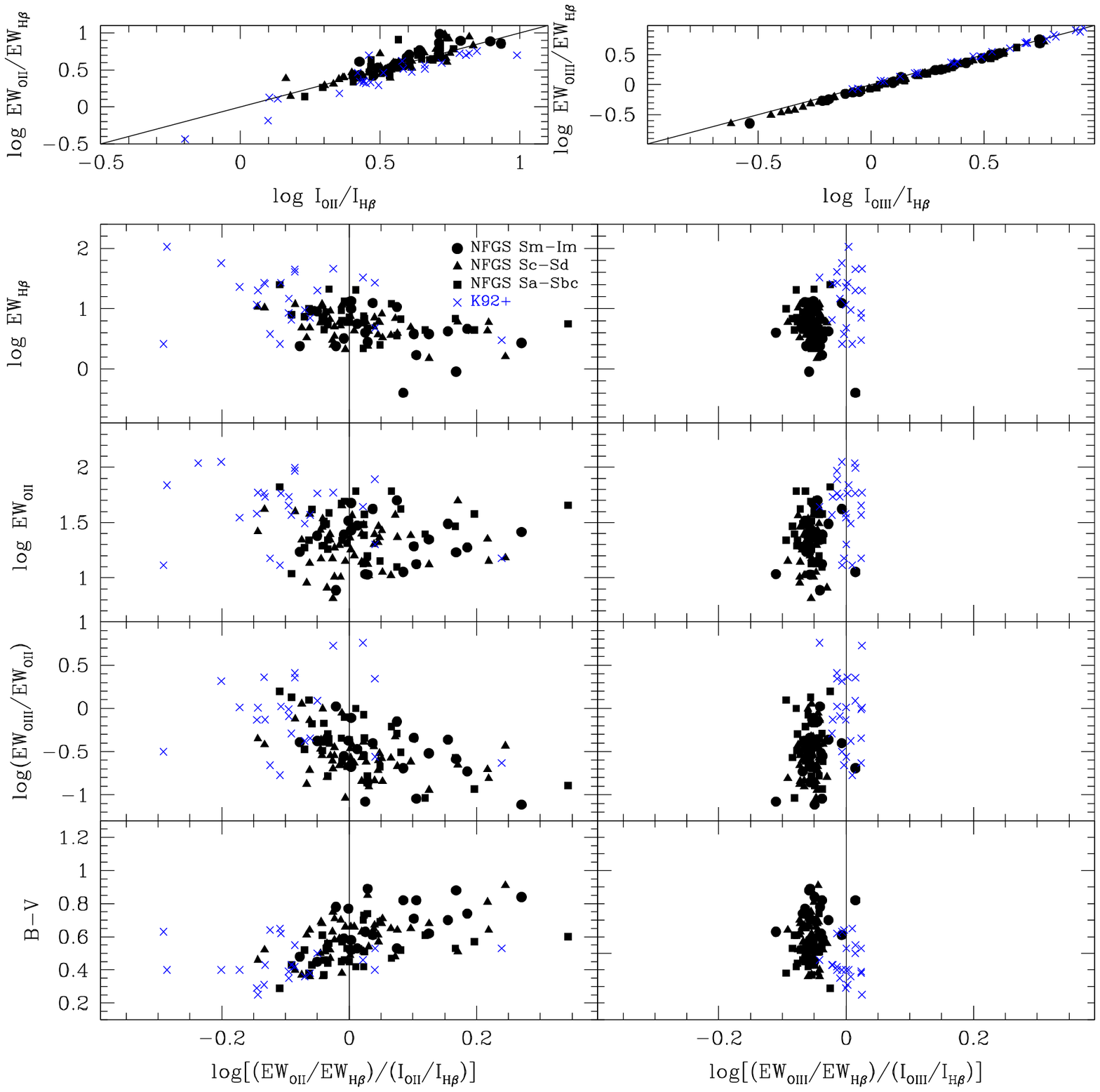}
\figcaption[EWtest3.cps] {Comparison of dereddened emission line fluxes 
and flux ratios
to emission line equivalent widths and EW ratios for galaxies from
Kennicutt (1992a,b) and Jansen \etal\ (2000a,b; NFGS).  Upper panels
show the [O~II]/H$\beta$ flux ratios and EW ratios with a line
illustrating the 1-to-1 correspondence.  Lower panels show residuals
from the 1-to-1 relation as a function of equivalent width and galaxy
color.  There is generally a strong correlation between flux ratios
and EW ratios.  Panels showing systematic residuals are discussed in
the text.  Note that the ratio $(\ew{\oii}/\ew{\hb}) / (I_{\oii}/I{\hb})$ is
a measure of $1/\alpha$.
\label{EWtest3} }
\end{figure}

Figure~\ref{EWtest3} compares the oxygen and hydrogen emission-line
flux ratios to equivalent width ratios as a function of emission line
strength and $B-V$ color for the K92+ and NFGS samples.  Solid symbols
denote the Nearby Field Galaxy Sample while crosses denote the K92+
galaxies.  The upper left panel compares the ratio of dereddened
[O~II] to $H\beta$ fluxes, $I_{\oii}/I_{H\beta}$, versus the ratio of
[O~II] to $H\beta$ equivalent widths, $\ew{\oii}/\ew{\hb}$.  A solid
line marks the 1-to-1 correspondence.  There is a good correlation
between the two quantities, indicating that strong-line equivalent
widths are a good surrogate for dereddened line fluxes. The RMS
deviation from the 1-to-1 correspondence is
$\sigma(\log[\ew{\oii}/\ew{\hb}]) = 0.11$ dex for the combined
K92+NFGS samples.  Similarly, the panel at top right shows the
$I_{\oiii}/I_{H\beta}$ versus $\ew{\oiii}/\ew{\hb}$ ratios and
indicates that these equivalent width ratios are an excellent
substitute for dereddened line fluxes, as expected.  The RMS
deviation from the 1-to-1 correspondence is
$\sigma(\log[\ew{\oiii}/\ew{\hb}]) = 0.05$ dex for the combined
K92+NFGS samples.

The lower panels of Figure~\ref{EWtest3} show residuals from the
1-to-1 line as functions of $\ew{\hb}$, $\ew{\oii}$, the ratio
$\ew{\oiii}/\ew{\oii}$ and galaxy color.  The K92+ and NFGS galaxies
have very small residuals in the right column which compares [O~III]
to $H\beta$ ratios. Although small, the NFGS residuals are slightly
systematic, with an offset of 0.07 dex.  The excellent correspondence
of [O~III] fluxes to equivalent widths may be easily understood since
the [O~III] $\lambda\lambda$4959,5007 lines are close in wavelength to
H$\beta$ so that neither changes in the underlying galaxy continuum
light nor relative extinction will alter this ratio.  The same data is
shown for the [O~II] to $H\beta$ ratios in the left column.  Here, the
K92+ and NFGS galaxies have small, slightly-systematic residuals,
particularly as a function of color.  In these cases, $\alpha$ is
unlikely to be quite near to unity.  There seems to be a systematic
offset between the NFGS and K92+ sample, possibly caused by a
systematic overestimate of the reddening correction to the [O~II]
$\lambda$3727 flux due to low spectral resolution in the K92 atlas.
Within the NFGS sample, the largest residuals are seen in the reddest
galaxies, consistent with expectations that the reddest galaxies will
have the smallest value of $\alpha$ (see Equation~\ref{EQionpar}).
The systematic trend seen in the residuals with respect to $\ew{\hb}$
is most likely a reflection of a correlation between $\ew{\hb}$ and
color, i.e., redder galaxies have relatively less star formation and
hence relatively weaker $\ew{\hb}$.  These same galaxies with low
$\ew{\hb}$ will also be the most affected by lack of correction for
underlying stellar absorption.  Nevertheless, the bulk of the
residuals are near zero, indicating that $\alpha$ has a typical value
near unity.

\begin{figure}
\plotone{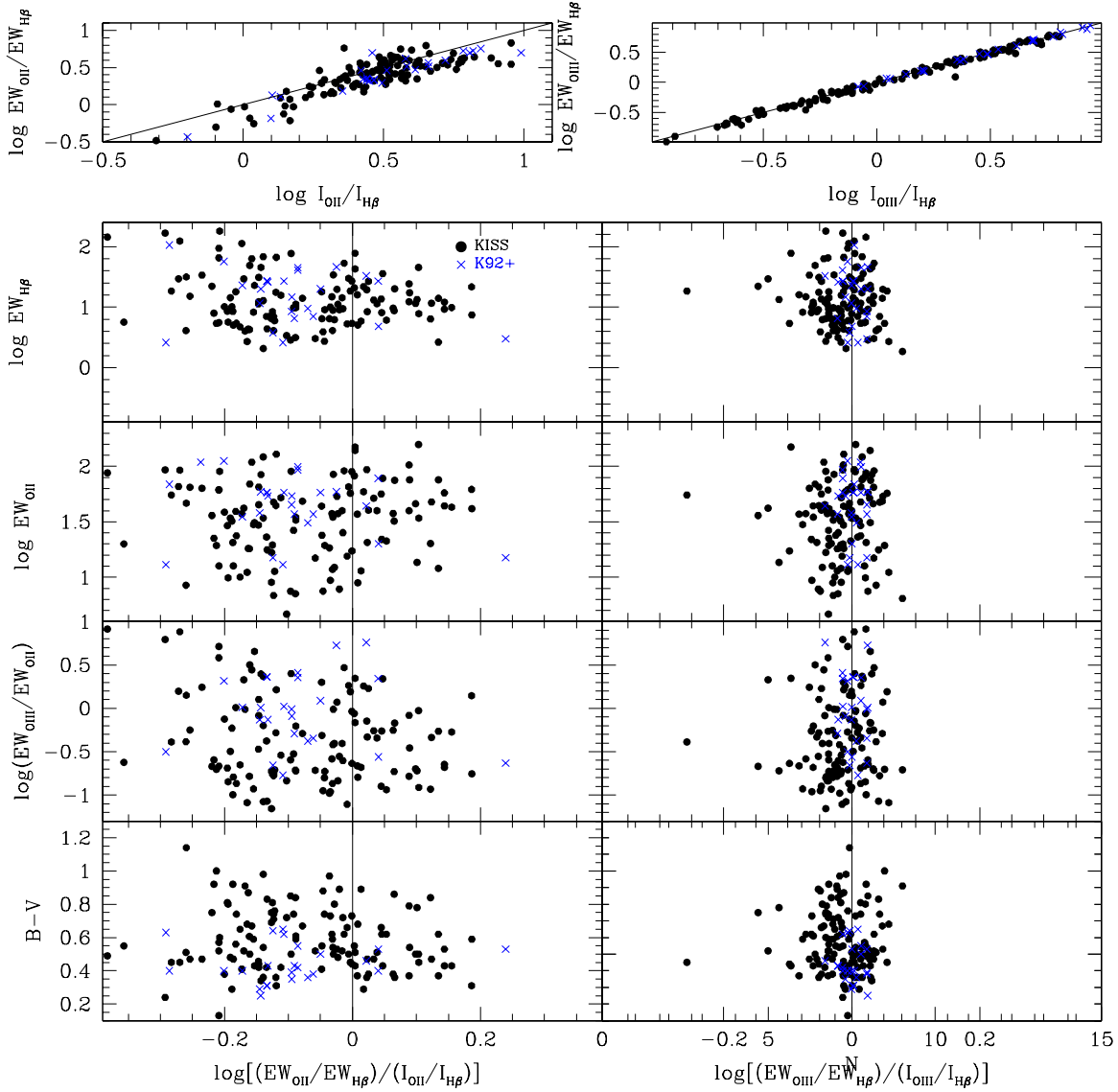}
\figcaption[EWtest3KISS.cps] {Comparison of dereddened emission line fluxes and 
flux ratios to emission line equivalent widths and EW ratios for
galaxies from the K92+ galaxies and the KISS (Salzer \etal\ 2001)
emission line galaxy
survey.  Upper panels show the [O~II]/H$\beta$ flux ratios and EW
ratios with the line illustrating the 1-to-1 correspondence.  Lower
panels show residuals from the 1-to-1 relation as a function of
equivalent width and galaxy color.  There is generally a strong
correlation between flux ratios and EW ratios.  Panels showing
systematic residuals are discussed in the text. 
\label{EWtest3KISS} }
\end{figure}

Figure~\ref{EWtest3KISS} shows the same comparison as
Figure~\ref{EWtest3} but now for the KISS and K92+ samples.  There
appear to be somewhat more scatter in the KISS sample than in
NFGS, and here the RMS deviation from the 1-to-1 correspondence is
$\sigma(\log[\ew{\oii}/\ew{\hb}]) = 0.15$ dex and
$\sigma(\log[\ew{\oiii}/\ew{\hb}]) = 0.04$ dex for the combined
K92+KISS samples.

The lower panels of Figure~\ref{EWtest3KISS} show residuals from the
1-to-1 line as a function of equivalent width and galaxy color. The right column
shows very small and non-systematic residuals, indicating excellent
correspondence between [O~III] equivalent widths and fluxes for all
KISS galaxy colors and line ratios.  However, the left column shows
significant dispersion of 0.15 dex between $\log[\ew{\oii}/\ew{\hb}]$
and $\log[I_{\oii}/I_{H\beta}]$.  The residuals are not correlated
with galaxy color or equivalent width, suggesting that measurement errors
and/or uncertainties in the reddening correction are responsible for the
dispersion.
The [O~II]/H$\beta$ residuals appear slightly offset from the zero line,
indicating $\alpha$ slightly above unity on the average.
 

\subsection{The Ionization Parameter [O~III]/[O~II]}

We consider next the ratio of the [O~II] and [O~III] lines, as this
was shown to be the most direct way to estimate the unknown value $\alpha$.
In addition, the quantity $I_{\oiii}/I_{\oii}$ is a measure of the
hardness'' of the ionizing photons, and is used in the $R_{23}$ to O/H
conversion.  Modern calibrations relating $R_{23}$ to oxygen abundance include a
measure of the ionization parameter, usually $O_{32}\equiv
(I_{4959}+I_{5007})/I_{3727} = I_{\oiii}/I_{\oii}$ as a second
parameter (e.g., McGaugh 1991; Pilyugin 2001).

\begin{figure}
\plotone{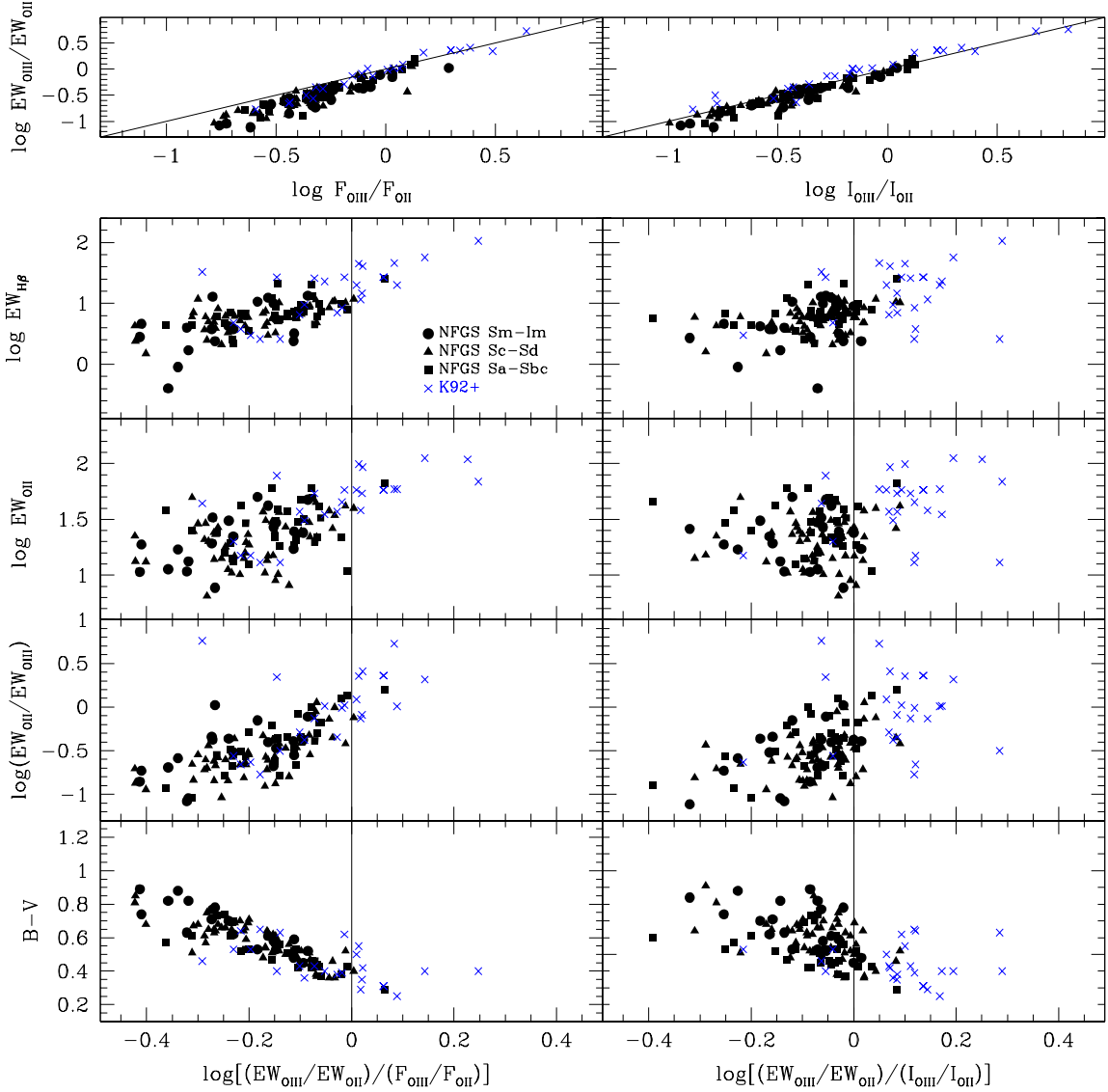}
\figcaption[EWtest5.cps] 
{Comparison of the ionization parameter quantity
$\ew{\oiii}/\ew{\oii}$ with $F_{\oiii}/F_{\oii}$ and
the dereddened ratio $I_{\oiii}/I_{\oii}$ for the K92+
and NFGS galaxies.  Lower panels show the residuals from the 1-to-1
correspondence as a function of line strength and galaxy color.
The tight correlation between $(\ew{\oiii}/\ew{\oii}) / (F_{\oiii}/F_{\oii})$
and $(B-V)$ is expected,
as the latter simply represents the ratio of spectroscopically
measured continuum flux under the lines.
The ratio $(\ew{\oiii}/\ew{\oii}) / (I_{\oiii}/I_{\oii})$ gives the
unknown value $\alpha$.
\label{EWtest5}  }
\end{figure}

In Figure~\ref{EWtest5}, we compare the equivalent width and flux
ratios for the K92+ galaxies (crosses) and NFGS galaxies (solid
symbols).  For completeness sake, on the left we plot
$\ew{\oiii}/\ew{\oii}$ against the observed $F_{\oiii}/F_{\oii}$
ratio; there is a clear correlation, but the points are scattered
mostly below the 1-to-1 correlation (shown by the line), consistent
with the lack of reddening correction.  Residuals to the 1-to-1 line
are shown in the lower panels, plotted against other galaxy parameters
The strongest correlation is seen against color, which is completely
expected: the ratio $(\ew{\oiii}/\ew{\oii}) / (F_{\oiii}/F_{\oii})$ is
identical to the ratio of the continua, $F_{C\lambda3727} /
F_{C\lambda4959,5007}$, which will be strongly correlated with $B-V$.
Most of the other systematic trends in the residuals can be explained
as correlations of the other parameters with $B-V$.

The right column shows a comparison of equivalent width ratios to the
dereddened line ratios.  This is of considerable interest, since the
ratio $(\ew{\oiii}/\ew{\oii}) / (I_{\oiii}/I_{\oii})$ is a direct
measure of $\alpha$ (see Equation~\ref{EQionpar}).  Here there is a
much tighter correlation, although there are still some systematic
trends in the residuals, again most strongly connected to color.  We
note, however, $\ew{\oiii}/\ew{\oii}$ is actually very close to
$I_{\oiii}/I_{\oii}$, which confirms that {\it $\alpha$ is near unity}
on the average.  We also note that there appears to be a systematic
offset between the NFGS and K92+ galaxies, perhaps related to the
offset seen in the NFGS [O~III] data (Figure~\ref{EWtest3}, middle
column).  While we might try to define a color-dependent term for
$\alpha$, it is unclear how much of the systematic trend is due to the
offset between samples.  Without any color-dependent term, the scatter
in $\alpha$ about unity is only $\sigma = 0.12$ dex.

\begin{figure}
\plotone{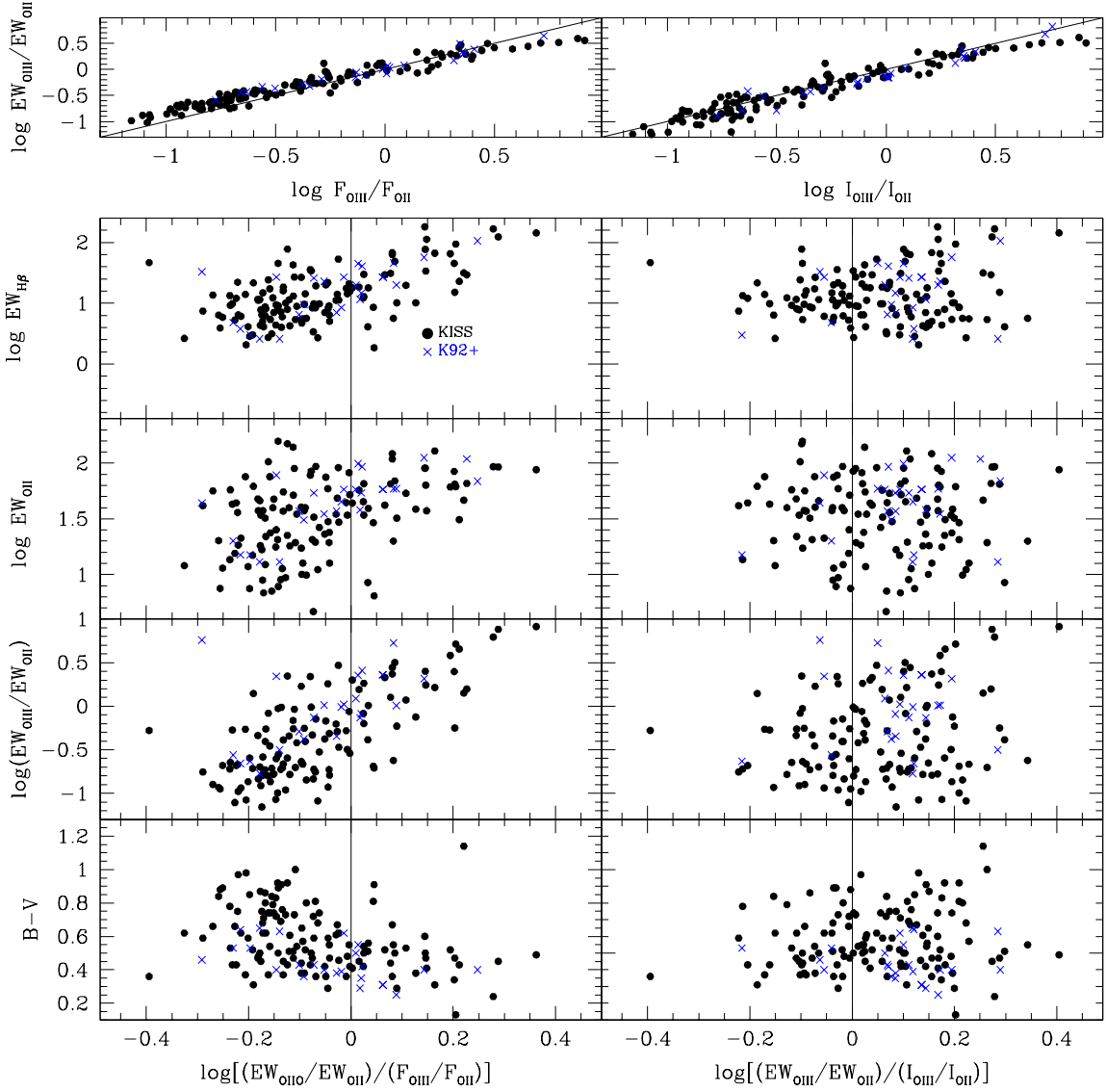}
\figcaption[EWtest5KISS.cps] 
{Same as Figure~\ref{EWtest5} except with KISS in place of NFGS galaxies.
The lack of a tight correlation between $(B-V)$ and
$(\ew{\oiii}/\ew{\oii}) / (F_{\oiii}/F_{\oii})$
is most likely due to observing different regions of each galaxy for spectroscopy
and photometry (see text).
\label{EWtest5KISS}  }
\end{figure}

Figure~\ref{EWtest5KISS} shows the same comparison for the KISS and
K92+ samples.  Here, the tight relationship between $B-V$ and the
residuals to $(\ew{\oiii}/\ew{\oii}) / (F_{\oiii}/F_{\oii})$ is
surprisingly lacking.  Since we are effectively comparing
$F_{C\lambda3727} / F_{C\lambda4959,5007}$, measured
spectroscopically, with the $B-V$ color determined from images, this
suggests considerable scatter in the continuum measurements.  Since
the KISS data span a redshift range up to $0 < z < 0.1$, it is
possible that some of the scatter is due to [O~III] contamination of
the $V$ broadband flux, but this would tend to reduce the $B-V$ colors
of those objects with high $\ew{\oiii}$ at higher $z$.  When we
correct for this in a reasonable fashion, we do not see a significant
reduction in the scatter, so we must assume the scatter due to some
other reason. The cause is almost certainly that the KISS spectroscopy
is not integrated across the galaxy, so the spectra tend to be
dominated by the nuclear regions, whereas the $B-V$ colors are from
the integrated galaxy light.

Looking at the dereddened line ratios in these data, we do not find
any trend in $\alpha$ with color, although such a trend might be
obscured by the scatter.  Nevertheless, the scatter about unity is
only $\sigma = 0.14$ dex, and again a value of $\alpha$ near unity is
indicated.

In summary, the local data empirically confirm that adopting $\alpha =
1$ is a reasonable approximation, with errors around 35\%, and
therefore the ratio $\ew{\oiii}/\ew{\oii}$ is a reasonable substitute
for the ionization parameter, $I_{\oiii}/I_{\oii}$.


\subsection{$EWR_{23}$ versus $R_{23}$}

\begin{figure}
\plotone{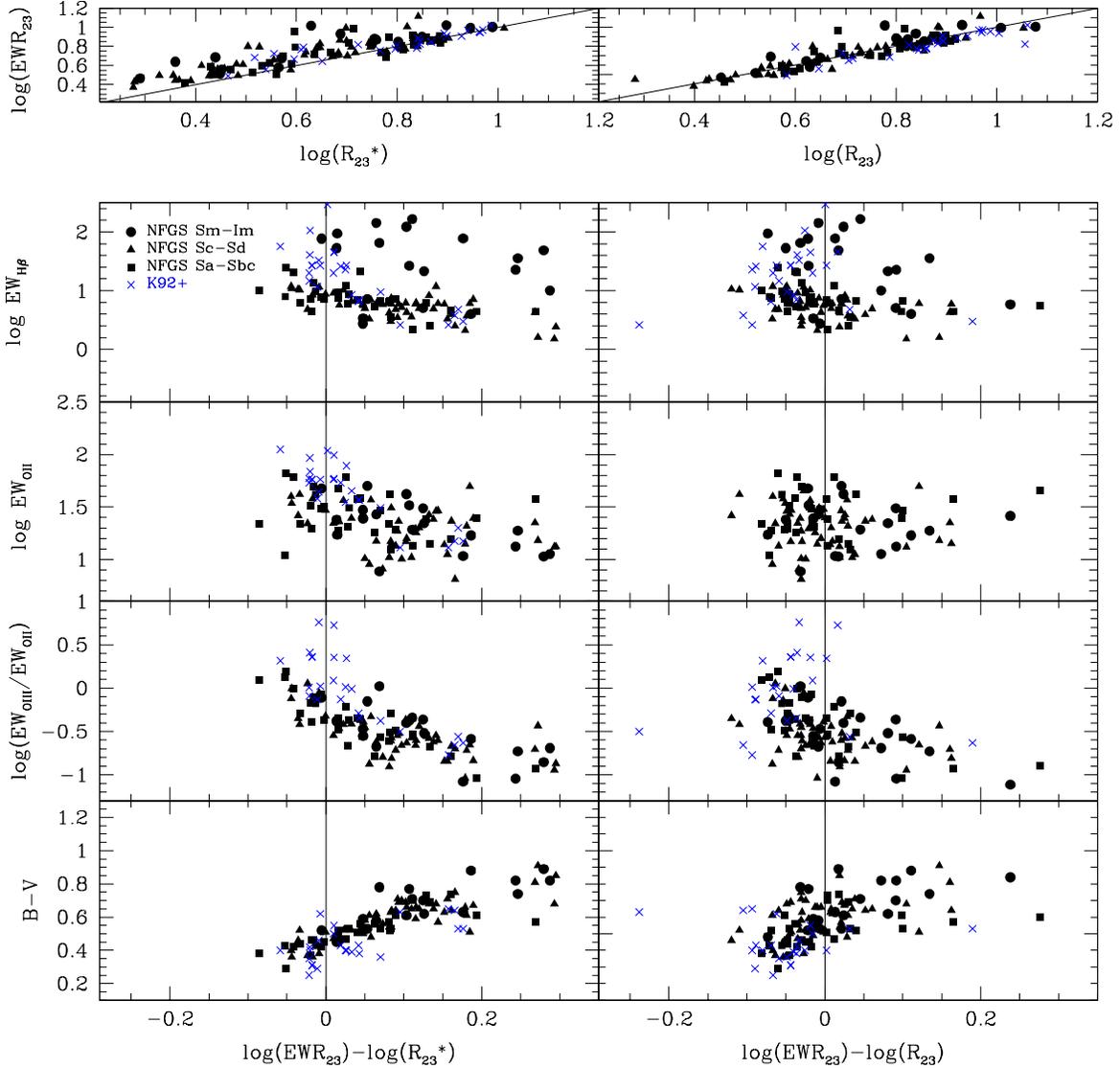}
\figcaption[EWtest2.ps] {
Comparison of the quantity $R_{23}$ and $R_{23}^*$ with $EWR_{23}$.
$R_{23}^*$ is $R_{23}$ without correction for reddening.  The strong
correlation between $R_{23}^*$ and $EWR_{23}$ (upper left panel) is
even stronger for $R_{23}$ and $EWR_{23}$, suggesting that 
oxygen abundances can be estimated from equivalent width ratios
as well as from dereddened line fluxes. 
The improvement of the correlation with $R_{23}$ shows that $EWR_{23}$
is successful in correcting at least part of the reddening.
The RMS dispersion from the 1-to-1 relation is
$\sigma(\log[R_{23}]) = 0.07$ dex.   Lower panels explore
the residuals in the correlation as a function of line strength, line
ratio, and galaxy color.  {\it left column}: no correction for
extinction or underlying Balmer absorption in either quantity; {\it
right column}: line fluxes have been corrected for reddening.
There remains a small systematic trend with color, but, excluding only
a few outliers, the $EWR_{23}$ and $R_{23}$ values agree to about $\pm 0.1$ dex
for this sample.
\label{EWtest} }
\end{figure}

The previous sections suggest that the equivalent width ratios can be
simply substituted into the $R_{23}$ formula, and here we provide
a direct test of that.  We define the measure
\begin{equation}
  EWR_{23} \equiv {{\ew{\oii\lambda3727} + \ew{\oiii\lambda4959} +
 	\ew{\oiii\lambda5007} }\over{\ew{H\beta}} }.
\end{equation}
which should be approximately equivalent to $R_{23}$.  For completeness,
we will compare $EWR_{23}$ against both the reddening-corrected $R_{23}$
and the analogous measure formed with non-corrected line fluxes, $R_{23}^*$.
Figure~\ref{EWtest} (upper left panel) shows the comparison
between $R_{23}^*$ and $EWR_{23}$ for K92+ and NFGS galaxies
constructed from the raw fluxes and equivalent widths.
A solid line illustrates a 1-to-1 correspondence.  The lower
rows in the left column of Figure~\ref{EWtest} show residuals from the
1-to-1 correspondence as a function of $\ew{\hb}$, $\ew{\oii}$,
$\ew{\oiii}/\ew{\oii}$ and galaxy $B-V$ color.  The correlation
between $R_{23}^*$ and $EWR_{23}$ is strong but has considerable
scatter.  Formally, the RMS dispersion from the 1-to-1 relation is
$\sigma(\log[R_{23}^*]) = 0.12$ dex for the combined K92+NFGS
samples.

The correlation is strongest for objects in the K92+ sample and for
objects with large values of $R_{23}^*$.  Deviations from 1-to-1 are
greatest for objects in the NFGS sample which have low $\ew{\hb}$,
high $\ew{\oiii}/\ew{\oii}$ ratios, and red colors.  These
systematic residuals may be understood as a consequence of the lack of
corrections for extinction.  [O~II] $\lambda$3727 is significantly
affected by extinction compared to the H$\beta$ and [O~III] lines.
The measured [O~II] flux is a lower limit to the true unextincted
intensity whereas the measured [O~II] equivalent width should be
unaffected by extinction provided that the extinction toward the gas
and stars are similar (but see Calzetti, Kinney \& Storchi-Bergmann 1994
for evidence that this assumption is sometimes invalid).
The strong systematic residuals with $B-V$ color seen in
the lower left panel is a most likely a consequence uncorrected
extinction, since redder galaxies are often those with greater
extinction.

In the right column of Figure~\ref{EWtest} we show a similar
comparison of $EWR_{23}$ with $R_{23}$, where $R_{23}$ has been
corrected for reddening using the theoretical Balmer decrement.  Here
the correlation is much stronger.  The strong correlation between
galaxy color and residuals seen in left column is now mostly gone,
suggesting that the [O~II] line fluxes have been successfully
corrected for reddening.  The RMS dispersion from the 1-to-1 relation is
$\sigma(\log[R_{23}]) = 0.07$ dex for the combined K92+NFGS samples.
The ratio of equivalent widths is a good substitute for the
reddening-corrected $R_{23}$ ratio.  Use of equivalent widths will
be superior to line ratios if the reddening corrections are not known,
as in the case of galaxies for which the H$\alpha$/H$\beta$ ratio is not
available (typically true for redshifts $z > 0.3$).
The residuals in the lower panels of column 2 are mostly symmetric about
zero, with the largest scatter again occurring for objects with very
low values of $\ew{\hb}$.  Some of the systematic residuals
are probably also caused by varying continuum shapes, especially
among the NFGS, which affect the equivalent widths of
the [O~II] lines in a systematic manner which is related to
galaxy color and the average age of the stellar population.
In any case, the RMS of 0.07 dex in
$R_{23}$ will often be on the same order as, or even less than the
statistical measurement uncertainties on the strong line equivalent
widths in high-redshift spectroscopic surveys, even when the signal-to-noise 
of the emission line equivalent widths is as low as 8:1.

\begin{figure}
\plotone{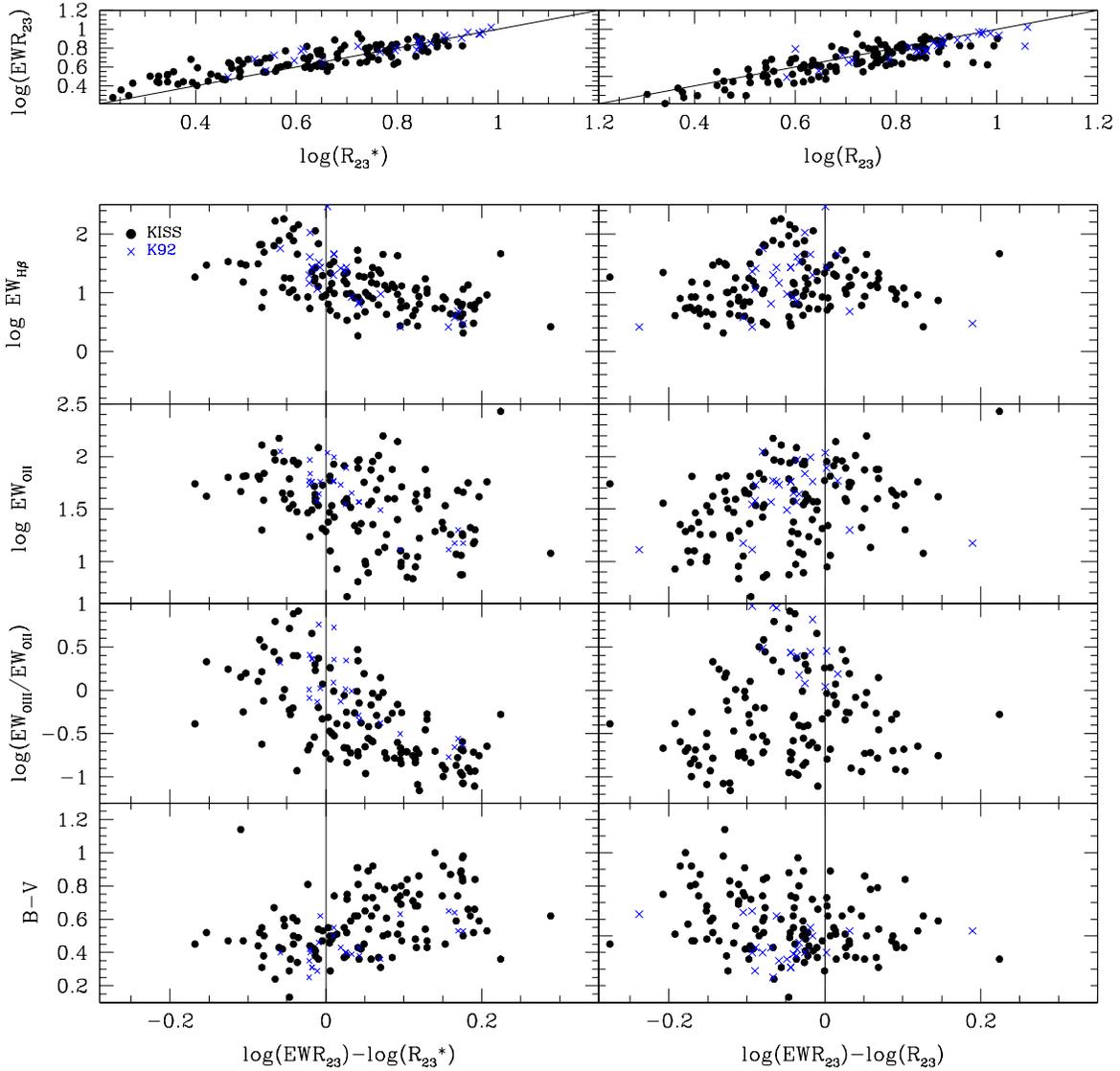}
\figcaption[EWtest2KISS.cps] 
{Comparison of the quantity $R_{23}$ and $R_{23}^*$ with $EWR_{23}$
for the K92+ and KISS galaxy samples. Upper panels show the
excellent correlation between $R_{23}$ and $R_{23}^*$ with $EWR_{23}$.
Lower panels show the residuals from the 1-to-1 correspondence as a
function of line strength and galaxy color.  Residuals are larger, but
less systematic, for the KISS galaxies than for the NFGS and K92+
galaxies. 
\label{EWtestKISS} }
\end{figure}

Figure~\ref{EWtestKISS} shows a comparison of $EWR_{23}$ with
$R_{23}^*$ and $R_{23}$ for the K92+ and KISS galaxy samples.  The
left column shows the comparison of $EWR_{23}$ with $R_{23}^*$ and the
associated residuals from the 1-to-1 correspondence.  There is a good
correlation between $EWR_{23}$ with $R_{23}^*$ in the upper left
panel.  The RMS dispersion from the 1-to-1 relation is
$\sigma(\log[R_{23}^*]) = 0.11$ dex for the combined K92+KISS
samples.  The left column shows systematic residuals with galaxy color
and line strength indicating that reddening is significant for some
galaxies.  The right column again shows the comparison of $EWR_{23}$ with
$R_{23}$ and the associated residuals as a function of equivalent width
and galaxy color.  The residuals for the KISS galaxies are now slightly smaller
and much less systematic after application of a reddening
correction. The RMS dispersion from the 1-to-1 relation is
$\sigma(\log[R_{23}]) = 0.09$ dex for the combined K92+KISS
samples.
The residuals are larger than for the K92+ and NFGS objects, and the KISS
galaxies do not show the systematics with galaxy
color or line strengths and ratios which the NFGS galaxies exhibit.
The lack of systematics with color is attributable to slit-vs-integrated light
issues, but this does not explain the larger scatter or lack of
systematics with pure line measures.
This latter is
possibly a result of the relative homogeneity of the KISS sample, as
KISS galaxies have stronger emission lines and do no include the
diversity of more quiescent galaxies with older stellar populations
found in the NFGS.

\subsubsection{Effects of Stellar Balmer Absorption}

Ideally, the quantity $R_{23}$ from which a metallicity is derived
should be computed using an $H\beta$ line strength which has been
corrected for both interstellar reddening {\it and} absorption by
atmospheres of the underlying stellar population.  In practice, the
amount of underlying absorption is difficult to measure even under
ideal circumstances with high signal-to-noise data.  Spectra of
distant galaxies frequently lack the signal-to-noise necessary to
measure multiple Balmer lines and correct simultaneously for
extinction and Balmer absorption in a self-consistent fashion.  For
galaxies with strong emission lines due to active star formation
(i.e., $\ew{\hb} >25$ \AA), a correction of a few \AA\ to the
$H\beta$ line will have a small impact on the derived $R_{23}$ or
$EWR_{23}$.  However, in galaxies dominated by older stellar
populations with weak emission lines, $R_{23}$ or $EWR_{23}$ will
depend sensitively on the correction for Balmer absorption.

Until this point in the analysis, we have not made any corrections for
stellar Balmer absorption.  The effect of underlying Balmer absorption
(specifically the amount of absorption in the H$\beta$ line,
$\ew{\hb}(abs)$ will depress the measured $F_{H\beta}$ and
$\ew{\hb}$.  This leads to systematically large $R_{23}$ or
$EWR_{23}$, and systematically low oxygen abundances for objects on
the upper (metal-rich) branch of the empirical calibrations.

\begin{figure}
\plotone{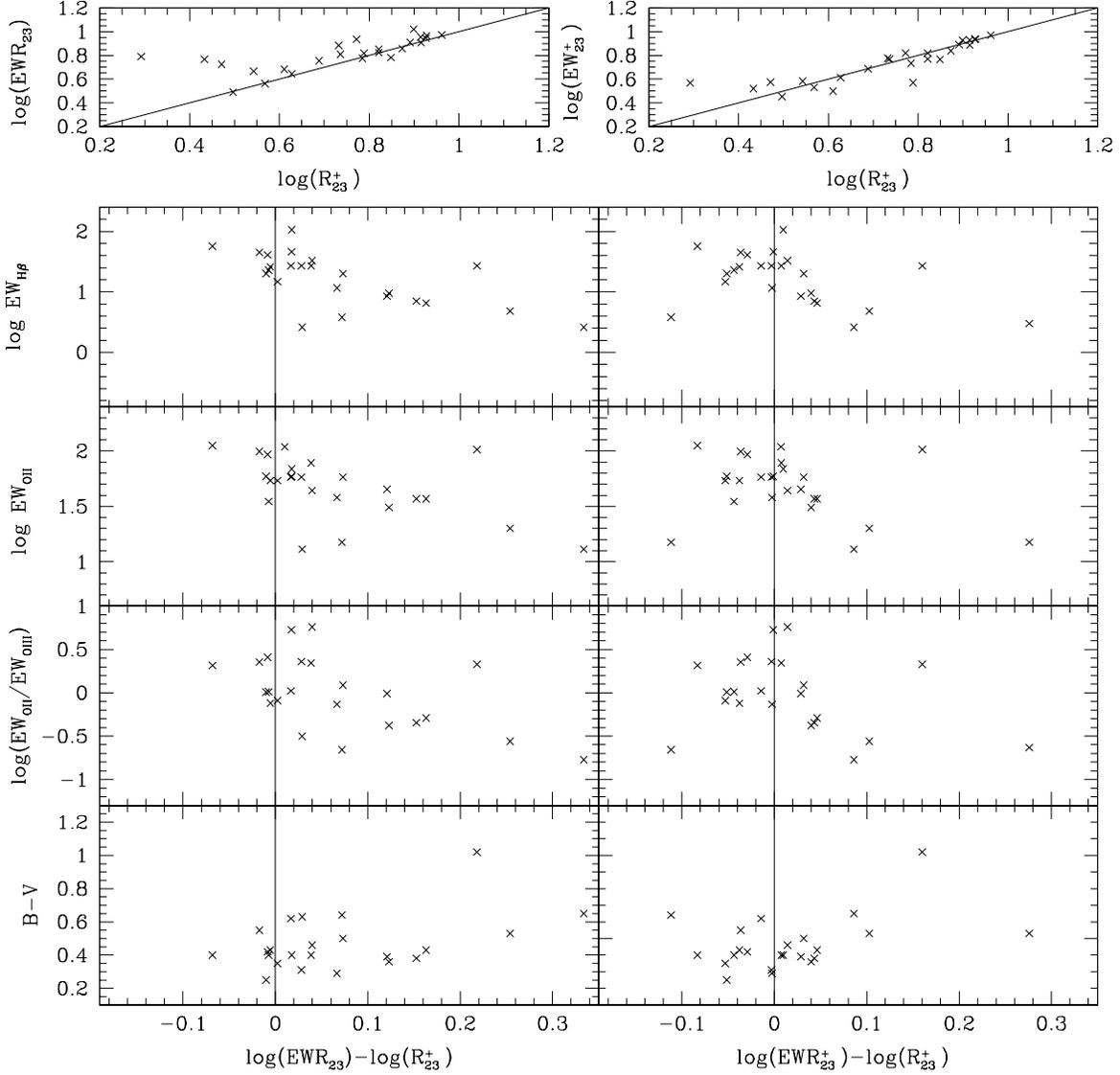}
\figcaption[EWtest4.ps] 
{Effects of correcting for underlying $H\beta$ absorption in K92+ galaxies.
The quantities $R_{23}^+ $ and $EWR_{23}^+ $ are analogous to $R_{23}$
and $EWR_{23}$ but with the $H\beta$ measurements corrected for stellar
absorption by 2 \AA, an average value found in a self-consistent fit
to both the reddening and stellar absorption using the available Balmer lines
(see text).
Upper panels show the correlation between $R_{23^+}$ and
with $EWR_{23}$ and $EWR_{23^+}$.  Lower panels show the residuals
from the 1-to-1 correspondence as a function of line strength and galaxy
color.  Galaxies with $\ew{\hb}\leq15 $ \AA\ are most seriously
affected by the lack of correction for stellar absorption.
\label{EWtest4} }
\end{figure}

The impact of stellar absorption can be assessed using
Figure~\ref{EWtest4}.  Using the K92+ galaxy spectra, we measured
several Balmer lines and performed a
self-consistent reddening and stellar absorption solution for each
galaxy (see Kobulnicky \& Skillman, 1996, for additional details).
The results indicated a range in underlying $H\beta$ absorption of
$\sim$1--3 \AA\ with an average value near 2 \AA.  From these
data we calculate a quantity $R_{23}^+$ which includes the corrections
for reddening and stellar absorption calculated for each galaxy.
The upper left panel of Figure~\ref{EWtest4} compares the ``raw'' $EWR_{23}$
ratio with the quantity $R_{23}^+$; the correlation is
modest, with a dispersion of $\sigma(R_{23}^+)=0.14$ dex and a
systematic offset of 0.06 dex.  The lower panels illustrate the nature
of the residuals as a function of galaxy color and line strengths.  As might be
expected, galaxies with the lowest $\ew{\hb}$ are the most deviant,
while galaxies with $\ew{\hb}>20$ show a much smaller dispersion.
A more logical approach is also to add a correction to $\ew{\hb}$ for the
underlying stellar absorption, forming a new quantity $EWR_{23}^+$.
A 2 \AA\ correction to $\ew{\hb}$ was chosen because it was
the mean correction needed to produce $R_{23}^+$, and is consistent
with mean corrections found for other galaxies (e.g., McCall \etal\
1985; Olofsson 1995).
The right top panel shows $EWR_{23}^+$ plotted against $R_{23}^+$.
The residuals are now much smaller with
$\sigma(R_{23}^+)=0.08$ dex and a systematic offset of less than 0.01
dex.
In the absence of direct measurements of the
Balmer absorption due to stellar populations, application of a 2 \AA\
blanket correction to $\ew{\hb}$ appears to be prudent for the types
of starforming galaxies in the K92+, DGSS and KISS samples.  
Samples including galaxies with older stellar populations may
require larger corrections.

\section{Discussion and Conclusions}

The ratios of the equivalent widths of strong oxygen and hydrogen
emission lines from the ionized component of distant galaxies can be
used as a measure of the global ISM metallicity via the substitution
of $EWR_{23}$ for $R_{23}$. We recommend the use of $EWR_{23}^+$ where
the $\ew{\hb}$ has been corrected for stellar Balmer absorption
assuming a mean correction of 2 \AA.  The typical dispersion from the
1-to-1 relation between either $EWR_{23}$ or $EWR_{23}^+$ and the
canonical reddening and absorption-corrected $R_{23}$ ratio is
$\sigma(R_{23})=0.08$ dex.  Residuals are somewhat smaller
($\sigma=0.05$ dex) for galaxies with the largest emission-line
equivalent widths (i.e., those having the largest rates of star
formation per unit luminosity).
The translation of these errors into errors in oxygen abundance will
depend on the value of $R_{23}$, but for the upper-metallicity
branch away from the ``turnaround'' point, the slope of log(O/H) vs
log($R_{23}$) is $\sim$1.4 (Pilyugin 2001), so the addition error
in log(O/H) would be $\sim$0.11 dex.
Thus, the additional uncertainty introduced
by translating a set of measured equivalent widths into the
traditional $R_{23}$ flux ratio diagnostic is comparable to or less
than the typical observational line measurement uncertainties and
systematic errors in the $R_{23}$ to O/H calibration which run 0.15
dex in O/H (e.g., Kobulnicky, Kennicutt \& Pizagno 1999 for a more
detailed discussion of the error budget).  We anticipate that the
method tested here will be useful for performing rough chemical
abundance estimates in large high-redshift galaxy samples. 
The approach described here
will be most useful in a statistical sense when large numbers of
objects are available for study.  Possible applications include
understanding the overall chemical evolution of star forming galaxies
over large intervals of cosmic time (Kobulnicky \etal\ 2003) or
assessing the impact of cluster environment and the intracluster
medium on the chemical properties of the ISM within galaxies (e.g.,
Skillman \etal\ 1996).

\acknowledgments

John Salzer helped make this paper possible by providing the KISS data
in tabulated form and by commenting on the manuscript.  
Detailed comments by an anonymous referee greatly improved the 
manuscript.  We thank Sheila Kannappan for a helpful discussion about the NFGS.
H.~A.~K.~was supported by NASA
through grant \#HF-01090.01-97A awarded by the Space Telescope Science
Institute which is operated by the Association of Universities for
Research in Astronomy, Inc. for NASA under contract NAS 5-26555.  This
work was also made possible by funding from the National Science
Foundation through grant AST-9529098, and NASA through
NRA-00-01-LTSA-052.

A.~C.~P.~would like to dedicate this paper to Gary Grasdalen, an early mentor,
with thanks.

\clearpage

\end{document}